\documentstyle[twocolumn,prb,aps,epsf]{revtex}
\begin{document}
\draft
\def \beq{\begin{equation}}
\def \eeq{\end{equation}}
\def \beqarr{\begin{eqnarray}}
\def \eeqarr{\end{eqnarray}}

\twocolumn[\hsize\textwidth\columnwidth\hsize\csname @twocolumnfalse\endcsname

\title{Inhomogeneous superconducting state in quasi-one-dimensional
systems
} 

\author{Kun Yang}

\address{
National High Magnetic Field Laboratory and Department of Physics,
Florida State University, Tallahassee, Florida 32306
}

\date{\today}
\maketitle

\begin{abstract}
We report on results of theoretical study of non-uniform
superconducting states in quasi-one-dimensional systems, with attractive 
interactions and Zeeman splitting between electron spins. Using bosonization
to treat intrachain electron-electron interactions, 
and a combination of 
renormalization group and mean-field approximation to tackle interchain
couplings, we obtain the phase diagram of the system, and show that the
transition between the uniform and non-uniform superconducting phases is a
continuous transition of the commensurate-incommensurate type.
\end{abstract}

]

The possibility of a superconducting state with inhomogeneous 
order parameter, stabilized by a sufficiently large Zeeman splitting between
electrons with opposite spin orientations due to either an external magnetic or
internal exchange field, was suggested more than thirty years ago by 
Fulde and Ferrell\cite{ff}
and Larkin and Ovchinnikov.\cite{lo}
Since then this Fulde-Ferrell-Larkin-Ovchinnikov (FFLO) state
has been the subject of a number of theoretical studies, but no direct
evidence of its existence has ever been found in conventional superconductors.
More recently it has attracted renewed interest in the context of organic,
heavy-fermion, and high-$T_c$ cuprate 
superconductors,\cite{gloos,yin,norman,rainer,shimahara,murthy,dupuis,modler,tachiki,geg,sh2,maki,samokhin,buzdin,yang,pickett,yang00,manalo,singleton,dan} 
as these new classes of superconductors are believed to provide conditions that
are 
favorable to the formation of
FFLO state due to their quasi-one or two-dimensionality as well as 
unconventional pairing symmetry. Indeed, some experimental evidence of its
existence has been reported.\cite{gloos,modler,geg,singleton}

The following picture emerged from the early theoretical studies (mostly of
mean-field type) of 
conventional s-wave superconductors subject to a Zeeman field $B$.
For sufficiently high field the system
is in the normal state. As the field strength decreases, at low temperatures
the system undergoes a
second-order phase transition at $B=B_{c2}(T)$
into the FFLO superconducting state. As the field
strength further decreases, another phase boundary is encountered at 
$B_{c1}(T)$,
and the system goes through another phase transition into the usual BCS 
superconducting state with uniform superconducting order parameter. 
While it is much more difficult to locate the position of $B_{c1}(T)$ than
$B_{c2}(T)$ (even in mean-field theory), 
as well as to address the nature of the transition there,
it has been widely
assumed\cite{gg} that this is a first-order phase boundary, across which the 
momentum of the order parameter and the magnetization
change discontinuously. 
This viewpoint was disputed in Ref. \onlinecite{rainer}, in which the authors
argue that the transition at $B_{c1}(T)$ is of second order. Thus the nature
of this transition is an unsettled issue.

In this paper we study quasi-one-dimensional (Q1D) superconductors subject to
a Zeeman field, and the possibility of formation of FFLO states in these
systems. Our motivation comes from two considerations. First of all, some of
the experimental candidates for FFLO state are made of weakly coupled
chains and therefore Q1D. Secondly, it is known in
that fluctuations are much stronger in low-dimensional systems than in 3D 
systems, and mean-field theories are much less reliable there.\cite{note}
On the other
hand the non-perturbative machinery developed for studying one-dimensional 
interacting electron systems (especially bosonization\cite{gogolin}) allows us
to go beyond 
mean-field theory and treat the intrachain electron-electron correlation
exactly in Q1D systems. 
In this paper we will take an approach that is similar to the one 
used in Ref.\onlinecite{carlson}, namely to treat the intrachain 
electron-electron interaction exactly using bosonization, and tackle the 
interchain couplings using a combination of renomalization group (RG) analysis
and mean-field approximation. Using this approach 
we are able to make a number of quantitative and reliable predictions 
about the FFLO state in these systems. In particular, we will show that 
the phase transition at $B_{c1}$ is {\em continuous} in these systems, 
and work out its
critical properties. For the sake of simplicity and concreteness, we restrict
our discussion to zero temperature throughout the paper.

We start by considering a one-dimensional electron gas with attractive 
interactions. In the bosonized form, the Hamiltonian reads
\begin{equation}
H=H_c+H_s+H_z,
\end{equation}
where $H_c$ and $H_s$ are the Hamiltonian for the charge and spin sectors
(which are decoupled, signaling the spin-charge separation):\cite{carlson}
\begin{equation}
H_\alpha=\int{dx}\left\{{v_\alpha\over 2}
\left[K_\alpha(\partial_x\theta_\alpha)^2
+{(\partial_x\phi_\alpha)^2\over K_\alpha}\right]
+V_\alpha\cos(\sqrt{8\pi}\phi_\alpha)\right\},
\end{equation}
where $\alpha=c$ or $s$, and $H_z$ is the
Zeeman coupling:
\begin{equation}
H_z=g\mu_BBS_z^{tot}={\sqrt{1\over 2\pi}}g\mu_BB\int{dx}\partial_x\phi_s(x).
\end{equation}
In these equations $\phi_c$ and $\phi_s$ are bosonic charge and spin fields
related to the (coarse-grained) charge and spin densities:
\begin{equation}
\rho(x)=\sqrt{2\over \pi}\partial_x\phi_c(x),
S_z(x)=\sqrt{1\over 2\pi}\partial_x\phi_s(x);
\end{equation}
while $\theta_\alpha$ are their dual fields satisfying
\begin{equation}
[\phi_\alpha(x), \partial_{x'}\theta_\alpha(x')]=i\delta(x-x').
\end{equation}

For attractive interactions, we typically have the Luttinger liquid parameters
$K_c > 1$ and $K_s < 1$ (for non-interacting electrons, we have $K_c=K_s=1$).
If the 1D electron gas is sufficiently far away from lattice commensuration, 
which we assume to be the case here, $V_c$ (which measures the strength of 
$4k_f$ Umklapp scattering) may be set to zero. Thus $H_c$ takes the form of
free massless bosons. On the other hand, in the spin sector $H_s$ has the form
of 1+1D quantum sine-don model, and for $K_s < 1$, $V_s$ (which measures 
the strength of back scattering between electrons with opposite spins) is 
relevant in the RG sense; at low energies it opens up a gap 
$\Delta_s\sim v_s\Lambda[V_s/v_s\Lambda^2]^{1/(2-2K_s)}$ ($\Lambda$ is the
ultraviolet cutoff)
for spin excitations.\cite{carlson} 
The elementary spin excitations are massive solitons (kinks and anti-kinks)
of the $\phi_s$ field, which carry spin $\pm 1/2$.\cite{kinknote} This spin gap 
$\Delta_s$ is the analog of quasi-particle gap in the BCS theory of higher
dimensional superconductors. The fundamental difference here, however, is that
in 1D there is {\em no} long-range superconducting order; instead the
correlation function of the Cooper pair operator decays with a power law.
The power law exponent can be calculated using the explicit representation of
electron operators in terms of boson fields:
\begin{equation}
\psi_{\lambda,\sigma}=N_\sigma\exp[i\lambda k_fx-i\Phi_{\lambda,\sigma}(x)],
\end{equation}
where $\lambda=\pm 1$ represents left/right movers, $\sigma=\pm 1$ represents
up/down spin particles, $N_\sigma$ is the Klein factor that also includes a
normalization constant, and
\begin{equation}
\Phi_{\lambda,\sigma}=\sqrt{\pi/2}[(\theta_c-\lambda\phi_c)+\sigma
(\theta_s-\lambda\phi_s)].
\end{equation}
Thus the singlet pair correlation function (at $T=0$)
\begin{eqnarray}
&&\langle\psi^\dagger_{+1+1}(x)\psi^\dagger_{-1-1}(x)\psi_{-1-1}(x')
\psi_{+1+1}(x')\rangle
\propto\nonumber\\
&&\langle\exp[i\sqrt{2\pi}(\theta_c(x)-\theta_c(x'))]\rangle
\langle\exp[i\sqrt{2\pi}(\phi_s(x)-\phi_s(x'))]\rangle
\nonumber\\
&&\propto|x-x'|^{-2\xi_{sc}},
\end{eqnarray}
where the scaling dimension 
\begin{equation}
\xi_{sc}={1\over 2K_c}.
\label{kc}
\end{equation}
Here we have used the fact that the spin field 
$\phi_s(x)$ is long-range ordered
in the spin-gapped phase.

Let us now consider the effect of $H_Z$. In $H_Z$ the Zeeman field couples to
the soliton density and plays the role of chemical potential of spin solitons.
In fact, $H_s+H_Z$ takes exactly the form of the Pokrovsky-Talapov 
model\cite{pt,schulz} 
which was introduced to study the two-dimensional classical
commensurate-incommensurate (CIC) transition. In our context, we thus 
expect a {\em continuous} CIC transition at
\begin{equation}
B=B_{c1}=2\Delta_s/g\mu_B,
\label{ci}
\end{equation}
beyond which spin solitons start to proliferate in the ground state.
Eq. \ref{ci} is an {\em exact} result because the Zeeman field couples to 
$S_z^{tot}$ which is a conserved quantity.\cite{sachdev}
In the incommensurate phase, the spin solitons form a spinless Luttinger
liquid with its own bosonized Hamiltonian, which describes the low-energy
spin excitations of the system:
\begin{equation}
H_{sol}=\int{dx}{v_{sol}\over 2}[K_{sol}(\partial_x\theta_{sol})^2
+(\partial_x\phi_{sol})^2/K_{sol}].
\end{equation}
In the long-wave length limit,
the soliton density field $\phi_{sol}(x)$ is related to the spin field 
$\phi_s$ through 
\begin{equation}
\phi_s(x)=\phi_{sol}(x)/\sqrt{2}+{\sqrt{\pi/2}}n_{sol}(B)x+{\rm const.},
\end{equation}
where $n_{sol}$ is the soliton density of the ground state.
In the limit $B\rightarrow B_c+0^+$, the solitons become extremely dilute and
the repulsive interaction among them become irrelevant; they can be treated as
spinless free fermions.\cite{schulz} As a consequence of this
we have (i) $K_{sol}=1$ and (ii) $n_{sol}(B)\propto (B-B_c)^{1/2}$ in this 
limit. Using these results we find in the incommensurate phase the 
superconducting correlation function 
\begin{eqnarray}
&&\langle\psi^\dagger_{+1+1}(x)\psi^\dagger_{-1-1}(x)\psi_{-1-1}(x')
\psi_{+1+1}(x')\rangle
\nonumber\\
&&\propto\exp[iQ(B)(x-x')]|x-x'|^{-2\xi'_{sc}}
\label{cisc}
\end{eqnarray}
where $Q(B)=\pi n_{sol}(B)$; approaching the phase boundary:
$B\rightarrow B_c+0^+$, we have
$Q(B)\propto (B-B_c)^{1/2}$ and
\begin{equation}
\xi'_{sc}={1\over 2K_c}+{1\over 4}=\xi_{sc}+{1\over 4}.
\end{equation}
This incommensurate phase (in the spin sector) is the 1D analog of the FFLO
phase in higher dimensional systems, as the appearance of the spin solitons
in the ground state induces an oscillatory phase in the superconducting 
correlation function, Eq. (\ref{cisc}). Also the ground state now has a finite
magnetization as in the FFLO phase, and the additional fluctuation due to
the soliton liquid makes the superconducting correlation function decays
faster in the incommensurate phase, in (loose) 
analogy to the fact that appearance
of unpaired quasiparticles reduces the size of the superconducting order
parameter in the FFLO phase. We emphasize again that here there is {\em no}
long-range superconducting order in either the commensurate or incommensurate
phases; also the CIC
transition is continuous as $n_{sol}$ increases continuously
from zero as $B$ crosses $B_{c1}$; both the magnetization and wave vector of
oscillation $Q$ are proportional to $n_{sol}$.

We now turn to the discussion of interchain couplings. The three leading 
potentially relevant perturbations to the decoupled Luttinger liquid fixed 
(dLL)
point are single electron hopping $H_{e}$, Cooper pair hopping (or Josephson
tunneling) $H_{J}$, 
and interchain $2k_f$
back scatterings $H_{C/SDW}$.\cite{dllnote}
For attractive interactions ($K_c > 1$),
$H_{C/SDW}$ is less relevant than $H_{J}$, in both the commensurate and
incommensurate phases. $H_{e}$ is irrelevant in the commensurate phase
due to the presence of a spin gap.
Since in this case the scaling dimension for $H_{J}$ is
\begin{equation}
\xi_J=2\xi_{sc}=1/K_c < 2,
\end{equation}
we conclude Cooper pair hopping is the leading relevant perturbation at the
dLL fixed point, and the system flows toward a superconducting 
phase with long-range superconducting order once interchain coupling is turned
on in the commensurate phase.

\begin{figure*}[h]
\centerline{\epsfxsize=9cm
\epsfbox{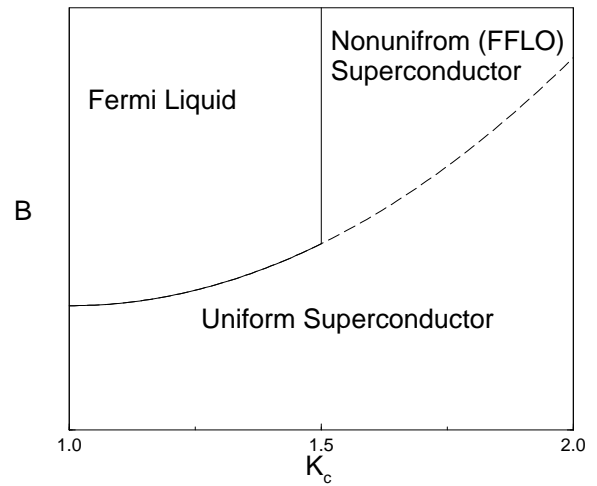}
}
\caption{
Schematic phase diagram of coupled Luttinger liquids subject to a Zeeman field.
The solid lines are first order phase boundaries while the dashed line is a
second-order phase boundary.
}
\label{phase}
\end{figure*}

Now let us consider the incommensurate phase. Right after the system enters
the incommensurate phase ($B\rightarrow B_{c1}+0^+$), we have 
\begin{equation}
\xi'_J=2\xi'_{sc}=1/K_c+1/2 < 2,
\end{equation}
thus $H_J$ is still relevant, albeit having a higher scaling dimension than that
in the commensurate phase. However in this case $H_{e}$ may also be relevant,
as there is no longer a spin gap in this case. We find in this case the
scaling dimension of $H_{e}$ to be
\begin{equation}
\xi'_e={1\over 4}(K_c+1/K_c)+{5\over 8}.
\end{equation}
We thus find that for $K_c > 3/2$, $\xi'_J < \xi'_e$, and $H_J$ is the leading
relevant perturbation at the dLL fixed point which drives the system to the 
Q1D
superconducting FFLO phase once interchain coupling is turned on. On the 
other hand, for $ 1< K_c < 3/2$, $\xi'_e < \xi'_J < 2$, and $H_e$ is the leading 
relevant perturbation at the dLL fixed point; in this case the system flows 
toward the high-dimensional Fermi-liquid fixed point.\cite{flnote} These
results are summarized in a schematic phase diagram, Fig. \ref{phase}. 
The phase boundary separating the Fermi liquid and the two superconducting 
phases are likely to be first-order since they are determined by the 
crossing of the scaling dimensions of two different relevant operators at the
dLL fixed point; on the other hand as we will argue below, the transition 
from uniform to FFLO superconducting phases is continuous. We emphasize in this
phase diagram we assume the Zeeman field $B$ is not too strong; if the 
Zeeman splitting is so strong as to be comparable to, say the Fermi energy, 
the continuum Luttinger liquid description of Q1D systems breaks down.

To address the nature of the transition between uniform and FFLO superconducting
phase, we focus on the pair hopping process and neglect other perturbations
that are less relevant:
\begin{eqnarray}
&H_J&=-\tilde{t}_J\sum_{\langle ij\rangle}
\int{dx}[\overline{\psi^i}_{+1+1}\overline{\psi^i}_{-1-1}
\psi^j_{-1-1}\psi^j_{+1+1}\nonumber\\
&+&\overline{\psi^i}_{+1-1}\overline{\psi^i}_{-1+1}\psi^j_{-1+1}\psi^j_{+1-1}
+h.c.]\nonumber\\
&=&-t_J\sum_{\langle ij\rangle}
\int{dx}\cos[\sqrt{2\pi}(\theta_c^i-\theta_c^j)]
\cos[\sqrt{2\pi}(\phi_s^i-\phi_s^j)],
\label{hj}
\end{eqnarray}
where $i$ and $j$ are chain indices, $\tilde{t}_J$ is the pair hopping
matrix element (or Josephson coupling strength), $\langle ij\rangle$ 
stands for 
neighboring chains, $t_J\propto \tilde{t}_J$,
and $h.c.$ stands for Hermitian conjugate.
In the case of decoupled Luttinger liquids, there is spin-charge separation 
and the CIC transition occurs in the spin sector. As we see in Eq. (\ref{hj}),
interchain pair hopping couples the spin and charge fields. On the other hand
since the system is in the superconducting phase (uniform or non-uniform)
in which the charge field $\theta_c$ is long-range ordered, in studying the
transition driven by $B$ we may use a mean-field approximation and replace
$\cos[\sqrt{2\pi}(\theta_c^i-\theta_c^j)]$ in Eq. (\ref{hj}) by its
expectation value: $\langle\cos[\sqrt{2\pi}(\theta_c^i-\theta_c^j)]\rangle
=C$. Clearly this expectation value depends on $B$ and it will also develop a
dependence on $x$ in the incommensurate phase; however as long as the dependence
is smooth across the transition (which would be the case if the transition is
continuous as we will show to be the case), we can treat it as a constant.
Thus in the mean-field approximation $H_J$ becomes
\begin{equation}
H_J^{MF}=-Ct_J\sum_{\langle ij\rangle}
\int{dx}\cos[\sqrt{2\pi}(\phi_s^i-\phi_s^j)].
\label{hjmf}
\end{equation}
Eq. (\ref{hjmf}) can also be obtained more formally by integrating out the
fluctuations of the $\theta_c$ field on top of its expectation value in the
Lagrangian formalism, which will yield a slightly renormalized coupling $C$.
The quantum Hamiltonian of $H_s+H_Z+H_J^{MF}$ can be mapped onto the problem
of classical CIC transition driven by $B$ at finite temperatures, 
in $d+1$ dimensions ($d$ is the physical dimension of the quantum problem
we study here). It is known that the CIC transition in higher dimensions is
still continuous, but the critical behavior is very different from the $d=1$
case considered earlier; in this case the density of domain walls (that consist
of solitons of individual chains aligned with true long-range order) depends
logarithmically on the distance from criticality:\cite{fisher}
\begin{equation}
n_{wall}\propto 1/\log(|B-B_c|^{-1})
\end{equation}
as $B\rightarrow B_c+0^+$. The wave vector of the inhomogeneous superconducting
order parameter $Q$ and the magnetization are both proportional to $n_{wall}$
and thus have the same dependence on $B$ near criticality.
We note that while we obtained these results by making a mean-field 
approximation to the (long-range ordered) charge fields, the main conclusion
that the transition is continuous should be robust; this follows simply from
the fact that the domain walls (whose appearance drives the transition) repel
each other, which is clearly the case here. The logarithmic dependence of 
$n_{wall}$ on $B-B_c$
then follows from the exponentially weak repulsion between the
domain walls. These in turn justify the
validity of the mean-field approximation employed.

To summarize, we studied formation of non-uniform superconducting state in
quasi-one-dimensional systems. Among our results include a phase diagram in 
terms of the Zeeman field and Luttinger liquid parameter. We also showed that
the transition between the uniform and non-uniform superconducting states is
continuous.

The author has benefited greatly from discussions with David Huse.
This work was supported by NSF DMR-9971541 and the A. P. Sloan Foundation.

\begin{references}
\bibitem{flnote} In principle there may be residual attractive interaction 
among quasiparticles that can destabilize the Fermi liquid fixed point that
the system flows toward, and trigger a superconducting instability. An analysis
of such possibilities is beyond the scope of the approach used here since we
can only analyze the possible instabilities of the dLL fixed point in this
approach.
\end{references}
\end{document}